\documentclass[fleqn,useAMS,usenatbib]{mn2e}

\bibliography{biblio}

\usepackage{mathptmx}

\usepackage{color,graphicx,amsmath,url}

\title[Number counts at 100\,$\mu$m, free from cosmic variance]{Extragalactic number counts at 100\,$\mu$m, free from cosmic variance}

\author[B.~Sibthorpe et al.]
{\parbox{\textwidth}{B.~Sibthorpe,$^{\! 1}$\thanks{E-mail: \texttt{bsibthorpe@gmail.com}}
R.\,J.~Ivison,$^{\! 1,2}$
R.\,J.~Massey,$^{\! 3}$
I.\,G.~Roseboom,$^{\! 2}$
P.\,P.~van der Werf,$^{\! 4}$
B.\,C.~Matthews$^{5}$ and
J.\,S.~Greaves$^{6}$
}
\vspace{0.4cm}\\
\parbox{\textwidth}{
  $^{1}$UK Astronomy Technology Centre, Royal Observatory Edinburgh,
  Blackford Hill, Edinburgh EH9 3HJ\\
  $^{2}$Institute for Astronomy, University of Edinburgh, Royal
  Observatory, Blackford Hill, Edinburgh EH9 3HJ\\
  $^{3}$Institute for Computational Cosmology, Durham University,
  South Road, Durham DH1 3LE\\
  $^{4}$Leiden Observatory, Leiden University, P.O.\ Box 9513, NL-2300
  RA Leiden, The Netherlands\\
  $^{5}$National Research Council of Canada, 5071 West Saanich Rd,
  Victoria, BC V9E 2E7, Canada\\
  $^{6}$School of Physics and Astronomy, University of St Andrews,
  North Haugh, St Andrews, Fife KY16 9SS}}

\begin{document}

\date{Accepted YYYY MMMM DD. Received YYYY MMMM DD; in original form YYYY MMMM DD}

\pagerange{\pageref{firstpage}--\pageref{lastpage}} \pubyear{2011}

\maketitle

\label{firstpage}

\begin{abstract}
  We use data from the DEBRIS survey, taken at 100\,$\mu$m with the
  PACS instrument on board the \emph{Herschel Space Observatory}, to
  make a cosmic variance-independent measurement of the extragalactic
  number counts.  These data consist of 323 small area mapping
  observations performed uniformly across the sky, and thus represent
  a sparse sampling of the astronomical sky with an effective coverage
  of $\sim$2.5\,deg$^2$.

  We find our cosmic variance independent analysis to be consistent
  with previous counts measurements made using relatively small area
  surveys.  Furthermore, we find no statistically significant cosmic
  variance on any scale within the errors of our data.  Finally, we
  interpret these results to estimate the probability of galaxy source
  confusion in the study of debris discs.
\end{abstract}


\section{Introduction}

One of the most fundamental measurements that can be made from
extragalactic survey data is that of source number counts.  In most
cases a balance between survey area and depth must be struck,
typically resulting in a small number of relatively deep compact
fields being observed alongside some wider shallow imaging.  For
example, the largest area surveyed using
\emph{Herschel}\footnote{Herschel is an ESA Space Observatory with
  science instruments provided by European-led Principal Investigator
  consortia and with important participation from NASA.}
\citep{Pilbratt2010} was the Astrophysical Terahertz Large Area Survey
\citep[H-ATLAS;][]{Eales2010}, which covers 550\,deg$^{2}$ but still
constitutes only 1.3 per cent of the entire sky.  In addition, the
H-ATLAS observations are relatively shallow, with deeper surveys
covering increasingly smaller areas \citep[e.g.][]{Elbaz2011}.
Consequently almost all source count measurements are potentially
subject to cosmic variance: the statistical uncertainty inherent when
inferring results based on data from a finite sub-region or
sub-regions of the sky.  Only by observing the entire sky can
  region-to-region fluctuations be averaged out entirely, giving a
  measurement truly free from cosmic variance.

In this letter we use multiple small-field observations
($\sim$28\,arcmin$^{2}$ each) distributed randomly across the entire
sky to obtain a cosmic variance independent assessment of the
extragalactic source number counts, and characterise the cosmic
variance on a wide range of angular scales.  This is achieved by
treating each field as representative of the larger region which it
inhabits. By calculating the source counts for all of these fields at
once, we can obtain a measurement which is representative of the
entire sky, 
in which cosmic variance is dramatically reduced compared to
contiguous area surveys of a similar size. By producing source counts
from various sub-combinations of these fields, we can also
estimate the cosmic variance -- quantitatively -- as a function of
angular scale, thereby determining the scale at which a survey can be
regarded as effectively free from cosmic variance.

For this analysis we use the 100-$\mu$m data from the Disc Emission
via a Bias-free Reconnaissance in the Infrared/Submillimetre survey
(DEBRIS; Matthews et al., in preparation), obtained with the
Photoconductor Array Camera and Spectrometer instrument \citep[PACS
--][]{Poglitsch2010} on board \emph{Herschel}.  We do not include an
analysis of the 160-$\mu$m DEBRIS data, obtained at the same time as
the 100-$\mu$m data, as their quality across the fully mapped area was
insufficient to achieve a useful measurement of the source counts.

This analysis is not intended to supersede the results in this band
from the PACS Extragalactic Probe \citep[PEP --][]{Lutz2011} presented
by \cite{Berta2010}, who use significantly deeper observations, or the
H-ATLAS results given by \cite{Rigby2011} and Ibar et al.\ (in
preparation), who use a significantly larger total area.  Instead this
analysis focuses on the comparison of these results with those
obtained from our cosmic variance independent measurement of the
source counts.  Various clustering analyses have already been
  performed at \emph{Herschel} wavelengths \citep[e.g.][]{Cooray2010,
    Maddox2010, Magliocchetti2011}, however, these have all been
  restricted to relatively small scales, with the largest clustering
  analysis at 100\,$\mu$m being at angular scales $\leq 6$\,arcmin.
  Consequently we aim to provide a complementary analysis, focusing on
  much larger angular scales.

In \S\,\ref{theData} we describe the data from which we obtain our
source counts, along with our reduction method.
\S\,\ref{catalogueSimulations} contains a description of our source
extraction method, and the simulations we performed to quantify
completeness.  Our number counts and measurements of cosmic variance
as a function of angular scale are given in \S\,\ref{results}.
These results are then interpreted to assess the impact of
  background source confusion on general debris disc surveys in
  \S\,\ref{debrisdiscs}.  Finally, our conclusions are given in
\S\,\ref{conclusions}.


\section{The data}\label{theData}

The DEBRIS survey is a \emph{Herschel} Open Time Key Programme whose
primary science goal is to discover and study debris discs around
nearby stars (Matthews et al., in preparation).  The target list is
generally unbiased, with sources only being excluded if the cirrus
confusion noise level at 100\,$\mu$m was predicted to be greater than
1.2\,mJy for a point source at the time of survey design.  The target
selection was taken from the Unbiased Nearby Star catalogue
\citep{Phillips2010} with sources being selected sequentially in order
of distance from the Sun.  With the exception of regions close to the
Galactic plane where cirrus confusion is high, the observed fields are
distributed uniformly on the sky, with no particular preference in any
one direction.  The flux limited observing strategy employed by the
DEBRIS team also ensures that each region is observed to the same
depth, thereby providing a uniform data set.

In this work we search for chance detections of extragalactic sources
in the extended imaging around 323 debris disc target fields.
The observations were made using the PACS `mini-scan map' observing
mode, which provides maps with an area of approximately 3.5\,arcmin
$\times$ 8\,arcmin, giving a total survey area of approximately
2.5\,deg$^{2}$.  The telescope scanning rate was
  20\,arcsec\,s$^{-1}$, and the full width half maximum of the
  resultant 100-$\mu$m beam was 7\,arcsec. This observing mode was
designed to observe compact and point-like sources, with the majority
of the integration time concentrated in the centre of the map.

The maps were reduced using the {\it Herschel} Interactive Pipeline
Environment \citep[HIPE;][]{Ott2010} version 7.0.  The standard
processing steps were followed and maps were made using the
`photProject' task. The data were high-pass filtered using a filter
scale of 16 frames, equivalent to 66\,arcsec., to suppress $1/f$
noise.  The final maps have a 1-$\sigma$ noise level of
$\sim$2.0\,mJy\,beam$^{-1}$ in the central region.

\section{Source catalogue and simulations}\label{catalogueSimulations}

Before finding and extracting sources we adjusted the maps to remove
any pixels whose integration time was below 25 per cent of the peak
map coverage.  This was done to remove the noisy edge regions which
would confuse a source-extraction algorithm.  A circular mask with a
radius of 45\,arcsec was also applied at the map centre to blank out
the original target star coordinates for each field.

We identified sources in the maps using the source-extraction
software, SExtractor \citep{Bertin1996}, and measured the flux density
using aperture photometry.  An aperture radius of 20\,arcsec was used
for all sources.  Custom aperture corrections were derived for this
dataset with reference to the aperture corrections provided by the
PACS instrument team.  These custom corrections were applied to each
measured source to obtain a final flux density measurement.

The fragmented nature of this survey and the highly variable,
irregularly shaped coverage within individual maps means it is
difficult to determine an accurate survey area and general noise level
for this work.  To overcome this issue we chose to adopt a reference
effective area with a regular shape for each field.  The reference
area is required to be larger than the range of the measured data and
in practice this was simply taken as the total area of the image,
including all regions for which no data were obtained.  A completeness
correction could then be determined based on simulations performed
within this reference area.  As the reference area is required
  to be sufficiently large to enable all position angles of the
  rectangular map, a significant portion of the reference area will
  always contain no data, meaning that the maximum completeness
  possible is $\sim20$ per cent.

To obtain the completeness correction for these data we placed ten
simulated point sources of a given flux density randomly into each
map.  We used observations of Arcturus (observation IDs
  1342188248 and 1342188249) -- obtained in the same observing
  configuration as our data -- as our point source model.  This model
  was scaled in order to return the expected flux density following
  application of the appropriate aperture correction.  No more than
  ten sources were input to a single map to avoid increasing the
  source density to a point wherein the simulated sources might blend
  and represent an increase in confusion
  noise\footnote{\url{http://herschel.esac.esa.int/Docs/HCNE/pdf/HCNE_releaseNote_v019_2.pdf}
  (nominally 0.1\,mJy\,beam$^{-1}$ at 100\,$\mu$m}). These simulated
sources were extracted using the same method as for our original
catalogue.  The number of simulated sources detected was then obtained
by differencing histograms constructed from the original and simulated
source catalogues.  The resultant histogram showed a Gaussian
distribution of sources centred at the flux density of the input
(simulated) source.  A Gaussian fit was made to this histogram and the
number of sources measured was compared to the number of simulated
sources to obtain the completeness at this flux density.  The fitted
width of the Gaussian was also used as a measure of the uncertainty in
flux-density measurement for all measured sources.  This
process was repeated for a range of flux densities to find the
completeness correction across the flux densities measured in our
original catalogue.  The entire simulation process was repeated five
times to reduce the uncertainty in the derived correction.

The completeness correction is a statistical result and is only valid
when working with the entire survey dataset and the adopted reference
map areas.  It is not applicable to specific individual regions of
real data within any given map. The final DEBRIS galaxy source
catalogue consists of 540 sources with a detection significance of
$\ge3\sigma$ and a typical 1-$\sigma$ uncertainty of $\sim3.0$\,mJy.
The typical completeness for this
  catalogue is 14 per cent, which equates to $\sim$65 per cent within
  the regions of the reference area (described above) in which data
  exist.  The large range of integration times within the map means
  that completeness levels near 100 per cent could only be achieved
  for the extremely rare, bright sources.

\section{Number counts and cosmic variance}\label{results}

\subsection{Number counts}

\begin{figure}
  \includegraphics[width=84mm]{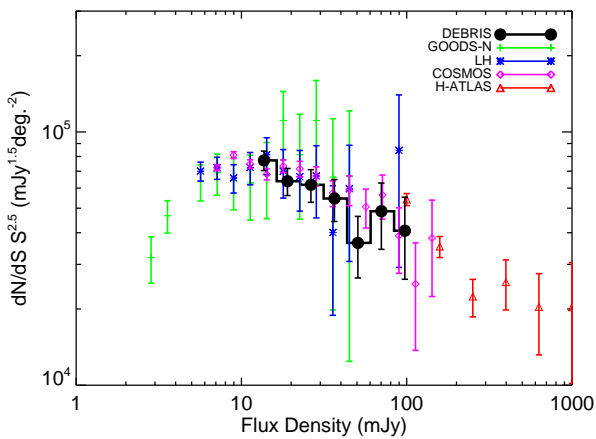}
  \caption{Galaxy source counts at 100\,$\mu$m, normalised to the
    Euclidean slope.  Source counts in various well-known survey
    fields from the PEP \citep{Berta2010} and H-ATLAS
    \citep{Rigby2011} surveys are overlaid for comparison.}
  \label{counts100um}
\end{figure}

The measured galaxy number counts at 100\,$\mu$m, normalised to the
Euclidean slope, are given in Table~\ref{table} and plotted in
Figure~\ref{counts100um}.  The counts obtained at the same wavelength
from the PEP \citep{Berta2010} and H-ATLAS \citep{Rigby2011} surveys
are also given, for comparison, with the three PEP counts being
obtained from 2-deg$^2$, 450-arcmin$^2$ and 140-arcmin$^2$ fields for
the Cosmic Evolution Survey \citep[COSMOS --][]{Scoville2007}, Lockman
Hole and Great Observatories Origins Deep Survey \citep[GOODS-N
--][]{Giavalisco2004} fields, respectively, and the H-ATLAS counts
from the 14-deg$^2$ science demonstration field.  Poisson noise
provides the dominant uncertainty in the DEBRIS number counts, with
the Eddington bias uncertainty ranging from $\sim$90 to
  $\sim$50 per cent of the corresponding Poisson noise level from low
  to high flux densities across the plotted range.  The errors quoted
  are the quadrature sum of these two sources of error.

\begin{table}
  \centering
  \caption{Measured extragalactic number counts, normalised to the Euclidean slope at 100\,$\mu$m}\label{table}
  \begin{tabular}{ccc}
    \hline
    Flux density & Number counts & 1-$\sigma$ error\\
    (mJy) & ($\times10^3$ mJy$^{1.5}$deg$^{-2}$) & ($\times10^3$ mJy$^{1.5}$deg$^{-2}$) \\
    \hline
    13.7  & 77.2     &  6.8 \\
    19.0  & 63.9     &  7.9 \\
    26.3  & 61.9     &  9.2 \\
    36.5  & 54.6     & 10.1 \\
    50.7  & 36.4     &  9.8 \\
    70.3  & 48.6     & 14.1 \\
    97.4  & 40.7     & 14.6 \\
    \hline
  \end{tabular}
\end{table}

Our number counts are in good agreement with those published by the
PEP and H-ATLAS teams, indicating that the results from these
individual fields are genuinely representative of the larger-scale
galaxy number counts within the given uncertainties of the relative
samples.

\subsection{Cosmic variance}

To measure cosmic variance on various angular scales we began with the
null hypothesis that there was zero cosmic variance on all scales
($\sigma^2=0$).  This hypothesis was then tested using the formalism
set out by \cite{Efstathiou1990}.  The sky was split into
circular cells of equal angular area, $A$, and galaxy number
counts, $N$, were obtained for each cell.  The cell-to-cell variations
were then compared with what would be expected from Poissonian
statistics using Equation 9 of \cite{Efstathiou1990},
\begin{equation}
\sigma^2 = \frac{\sum_j(N_j-A_j\sum_kN_k/\sum_kA_k)^2-(1-\sum_kA_k^2/(\sum_kA_k)^2)\sum_jN_j}{(\sum_jN_j/\sum_jA_j)^2[\sum_kA_k^2-2\sum_kA_k^3/\sum_kA_k+(\sum_kA_k^2)^2/(\sum_kA_k)^2]}.
\end{equation}
This generalised equation allows for incomplete cell sampling and is
therefore appropriate for the DEBRIS sample (see
\citeauthor{Efstathiou1990} for details).  The uncertainty in this
measurement ($\Delta(\sigma^2)$) was quantified using Equation 5 of
\cite{Efstathiou1990}.  This method was also implemented by
\citet{Austermann2010} to assess the level of cosmic variance in four
fields imaged with the AzTEC camera at 1.1\,mm.

\begin{figure}
  \includegraphics[width=84mm]{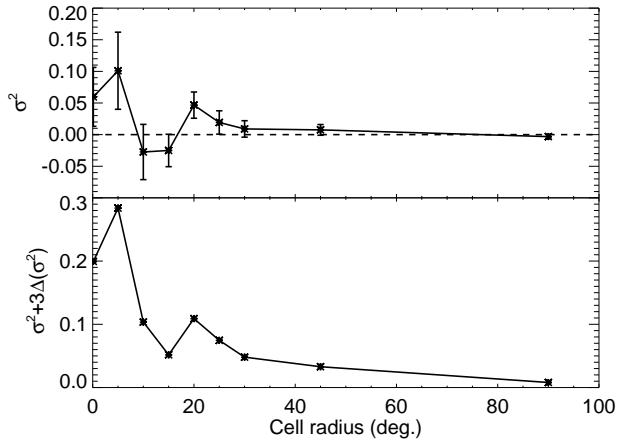}
  \caption{Measured cell variance (top) and upper limits (bottom) as a
    function of cell radius. Note that negative $\sigma^2$ indicates
    that the measured variance is less than that expected from
    Poisson statistics.} 
  \label{sig_sq}
\end{figure}

Cells with an angular area equal to that of one DEBRIS map up to
2$\pi$ steradians were tested.  The measured variance from each cell
size is shown in Figure~\ref{sig_sq}.  The measured variances have a
median value of $1.2\times10^{-2}$, in keeping with those reported by
\cite{Austermann2010} as well as the model of \cite{Moster2011},
within the stated errors. However, in all cases the signal-to-noise
ratio ($\sigma^2/\Delta(\sigma^2)$) is less than 3.  This indicates
that these measurements do not represent a statistically significant
detection of a non-zero cosmic variance on scales down to the
  size of one DEBRIS map.

Although the non-uniformity and sparse sampling of the DEBRIS
  data makes it difficult to calculate the angular correlation
  function for direct comparison with the results of
  \cite{Cooray2010}, \cite{Maddox2010} and \cite{Magliocchetti2011},
  it is still possible to compare their results with those from this
  analysis.  All three prior publications show a decrease in
  clustering (i.e.\ variance) with increasing angular scale.  The
  upper limits we obtain show a similar trend, albeit with very low
  statistical significance.

\cite{Cooray2010}, \cite{Maddox2010} and
  \cite{Magliocchetti2011} also find increased clustering with higher
  redshift galaxy populations.  Therefore our non-detection of cosmic
  variance at 100\,$\mu$m, which is typically dominated by low
  redshift ($z\le 1$) galaxies, is in keeping with these previous
  results.

The failure to detect cosmic variance in this work could be
  thought to be, in part, a result of the DEBRIS survey depth.  Deeper
  observations would probe higher-redshift galaxy populations, too
  faint to detect in the current data.  With deeper maps we would
  therefore expect to observe the same level of clustering at
  100\,$\mu$m as \cite{Cooray2010} and \cite{Maddox2010} find at 350
  and 500\,$\mu$m.  However, these deeper data would also contain
  additional faint, low-redshift galaxies with a similarly uniform
  distribution to the population of nearby galaxies currently
  observed, thereby diluting the clustering signature of these
  additional high-redshift sources.  In addition, if we assume that
  the clustering measured at 350\,$\mu$m is dominated by galaxies
  close to the 3-$\sigma$ confusion limit and at $z \sim 2.3$
  \citep{Cooray2010} then the required 1-$\sigma$ survey limit to
  detect these sources at 100\,$\mu$m would be $\sim$0.2\,mJy, only
  slightly higher than the 100-$\mu$m confusion limit.  Given these
  two factors, we conclude that our results are valid for all
  100-$\mu$m survey depths down to the limiting survey area of one
  DEBRIS map, given a comparable confusion noise limit.  Therefore, a
  survey of $\sim$28\,arcmin$^2$ would be expected to have $\sigma^2
  \leq 0.2$.

\section{Relevance to debris discs}\label{debrisdiscs}

Debris discs are most commonly identified by an infrared excess above
the predicted photospheric level.  It can be extremely difficult to
determine accurately whether a measured excess is due to a debris disc
or merely a chance alignment with a background galaxy. Some features
originally attributed to debris discs have subsequently been found to
be background objects, e.g.\ \cite{Greaves2005,Regibo2012}.  Confusion
is particularly problematic in statistical studies of debris discs. We expect
cases of background contamination will be more likely when analysing
many tens of discs, as is the case with the DEBRIS survey.  It is
therefore helpful to also interpret the results of this analysis in
terms of its relevance to debris disc studies.

In order to account for confusion, an accurate estimate of the
probability of a background source being present within a given radius
of a star is essential.  This study has shown that data from DEBRIS
observations, typical of almost all debris disc observations made with
\emph{Herschel}, are consistent with data from surveys of typical
extragalactic fields.  We have also shown that, to within the limits
of these data, there is no significant cosmic variance at 100\,$\mu$m.
Consequently, it is reasonable to make use of the results from deep
extragalactic surveys and apply their statistics directly to debris
disc observations.  For the purposes of estimating background source
confusion, we make the assumption that the invariance of the
100-$\mu$m counts can be extended to the other two PACS wavelength
bands.

Using the raw measured differential counts of \cite{Berta2011}, we
estimated the number of sources that could potentially explain an
excess at a given flux density.  To ensure all potential sources were
included, the number counts were weighted by a Gaussian probability
density function with a 1-$\sigma$ width equal to the uncertainty of
the original measurement.  This provides a measurement that includes a
contribution due to Eddington bias.  We calculated the number of
galaxies for all flux density bins given by \cite{Berta2011} over a
wide range of $\sigma$ values, then fitted a two-dimensional
polynomial to the output grid.  \cite{Magliocchetti2011} do
  find clustering in these data on small scales, but the level of
  clustering is extremely low, and none of the measurements is
  statistically significant ($<3\sigma$ in all cases).  As a result,
  we regard these data to be unclustered for the purposes described in
  this section.  The number of galaxies per square degree, $N$, at a
given flux density, $S_{\lambda}$, and measurement uncertainty,
$\sigma$, can thus be estimated directly from the resultant polynomial
fit,
\begin{equation}
\log_{10} N = \sum\limits_{i,j=0}^2 k(\lambda)_{i,j} \log_{10}\sigma^i \log_{10}S_{\lambda}^j,
\label{fitEqn}
\end{equation}

where $k(\lambda)_{i,j}$ is the matrix of polynomial coefficients
derived for the fit.  These matrices for flux densities measured in
mJy for each wavelength are,
\[
k(70\,\mu\rm{m}) = 
\begin{bmatrix} 
            3.34 &    -0.770 &    -0.470 \\
 -2.45\times10^{-4} &  -2.23\times10^{-3} &   5.23\times10^{-3} \\
  4.41\times10^{-6} & -2.31\times10^{-6} & -1.89\times10^{-5}
\end{bmatrix},
\]
\[
k(100\,\mu\rm{m}) = 
\begin{bmatrix} 
      3.84 &     -1.10 &   -0.181 \\
   -0.0108 &    0.0340 &  -0.0180 \\
  7.94\times10^{-5} & -2.58\times10^{-4} & 1.44\times10^{-4}
\end{bmatrix},
\]
\[
k(160\,\mu\rm{m}) = 
\begin{bmatrix} 
      4.12 &     -1.04 &   -0.142 \\
   -0.0208 &    0.0455 &  -0.0210 \\
  1.00\times10^{-4} & -2.35\times10^{-4} & 1.17\times10^{-4}
\end{bmatrix}.
\]
Therefore, given an expected background source number density $N(S_{\lambda},\sigma)$, the Poisson probability of $n$ sources existing within radius $r$ (in arcsec) is given by,
\begin{equation}
P(n,<r,N) = \exp\left(-\frac{N \pi r^2}{3600^2}\right) \frac{(N \pi r^2 / 3600^2)^n}{n!}.
\label{probEqn}
\end{equation}
Using this formalism it is possible to calculate the probability of a
background source being present in PACS observations of debris discs
with an excess ranging from $\sim$2--140\,mJy.  Table~\ref{probTable}
gives the probability of a source being present within a beam
half-width half-maximum radius of a given source location, for a range
of typical levels of measured excess flux density.  In all cases the
source is assumed to have been detected at the 3-$\sigma$ level.
These values have been calculated using Equation \ref{fitEqn} and have
been found to have a 1-$\sigma$ variation of 2.6$\times10^{-3}$ with
respect to probabilities calculated directly from the original data.

These results are suitable when estimating the confusion of an
individual source with a clearly measured excess.  However, when
working with a survey consisting of many observations, looking at the
probability of confusion at a specific flux density is less useful
than asking the question {\it what fraction of sources are likely to
  be confused by any number of background sources?}

\begin{table}
\centering
\caption{Probability, calculated using Equation \ref{fitEqn}, of one background source existing within a beam half-width half-maximum radius of the measured source loaction for a range of measured exccess flux densities.  In all cases the excess emission is assumed to be detected at the 3-$\sigma$ level.}\label{probTable}
\begin{tabular}{cccc}
\hline
$S_{\lambda}$ (mJy) & P$_{70 \mu\rm{m}}\times10^{-4}$ & P$_{100 \mu\rm{m}}\times10^{-4}$ & P$_{160 \mu\rm{m}}\times10^{-4}$ \\
\hline
3   &  11 &  59   & 300 \\
4   & 7.9 &  41   & 220 \\
6   & 4.5 &  24   & 130 \\
10  & 2.0 &  12   &  71 \\
20  & 0.6 &  4.2  &  28 \\
100 &     &  0.12 & 1.1 \\
\hline
\end{tabular}
\end{table}
The cumulative extragalactic number counts can be used to estimate the
probability of confusion, by one or more background sources, by
integrating the counts from the survey limiting flux density,
$S_{\rm{lim}}$, to the upper limit of the data.  This number of
galaxies, $N_{\rm c}$, can then be input to the cumulative Poisson
probability equation,
\begin{equation}
P_{\rm c}(n,<r,N_{\rm c}) = \exp\left(-\frac{N_{\rm c} \pi
    r^2}{3600^2}\right)\sum\limits_{i}^n \frac{(N_{\rm c}\pi r^2/3600^2)^i}{n!},
\end{equation}
where $P_{\rm c}$ is the probability of confusion with $n$ or more
background sources within a given radius $r$.

\begin{figure}
  \includegraphics[width=84mm,trim=5mm 5mm 5mm 5mm]{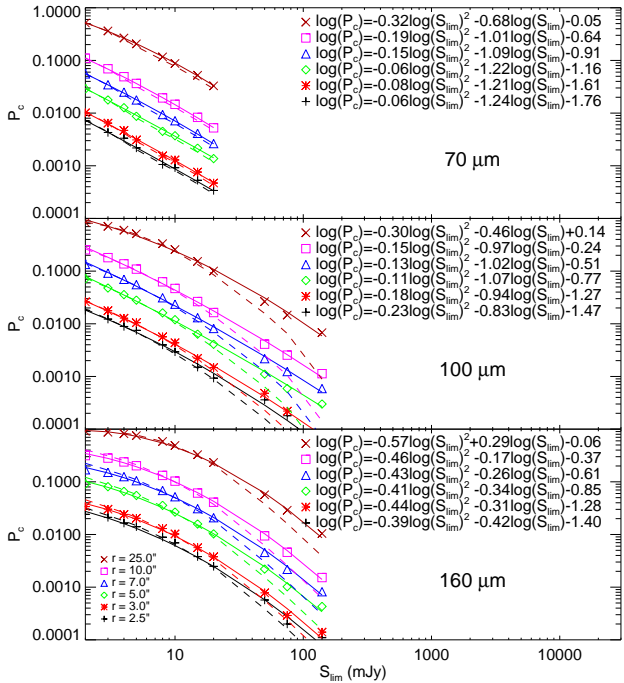}
  \caption{Probability of confusion by one or more background sources
    as a function of survey limiting flux density calculated via the
    Monte-Carlo method (symbols) and the na\"ive method (dashed line).
    Polynomial fits made in log-log space to the Monte-Carlo points
    are plotted as solid lines, with the fitted equations given in the
    top right of each panel.  In all cases the limiting flux density
    is assumed to be the 3-$\sigma$ limit for the given survey.
    Results are shown for confusion within various radii of a given
    source location.}
  \label{cumProb}
\end{figure}

Using the raw cumulative number counts, however, does not take account
of Eddington bias.  Due to the steep slope of the counts a larger
number of sources will be boosted above $S_{\rm{lim}}$ than will be
boosted down, making this `na\"ive' measurement of $N$ an
underestimate.  To assess the significance of Eddington bias in this
case we used the cumulative number counts from \cite{Berta2011} and
performed a Monte-Carlo simulation.  Simulations were performed for a
range of limiting survey flux density and region radii, and the output
is given in Figure~\ref{cumProb}.

Results from the na\"ive and Monte-Carlo methods are in good agreement
for faint survey limits, but begin to diverge at limits greater than
$\sim$10\,mJy, with the na\"ive method systematically underestimating
the confusion.  In these bright survey limit cases, the increased
measurement uncertainty results in a higher contribution from
Eddington bias and thus an underestimate in the na\"ive estimate of
$N_{\rm c}$.  Note that the increase in the measurement uncertainty is
a result of our assumption that the $S_{\rm{lim}}$ is a 3-$\sigma$
limit.  For higher survey thresholds, e.g. 5-$\sigma$, the
deviation between the two methods would begin at higher flux
densities.

These results are applicable for surveys without a single, limiting
flux density.  In this case the probability of confusion for each
observation, given its own limiting flux density, can be estimated and
combined for all observations in the standard way to estimate the
probability of source confusion.

\section{Conclusions}\label{conclusions}

We find our galaxy number counts -- based on observations that
sparsely sample the entire sky in a uniform manner -- to be in
agreement with previous studies whose results are derived from
relatively small-area surveys.  We also measure no statistically
significant non-zero cosmic variance on scales of $\sim$28\,arcmin$^2$
(the scale of a DEBRIS map) up to 2$\pi$ steradians, providing only
upper limits.  We conclude that traditional, relatively small-area
surveys, such as those undertaken by the PEP and H-ATLAS teams, are in
general sufficiently representative to measure and characterise the
extragalactic number counts in a universal context.


\label{lastpage}
\end{document}